\documentclass{article}
\usepackage[utf8]{inputenc}
\usepackage{amsmath,amssymb,pxfonts,accanthis}
\newtheorem{assumption}{Assumption}

\newtheorem{lem}{Lemma}
\newtheorem{thm}{Theorem}
\newtheorem{defn}{Definition}
\newtheorem{rem}{Remark}

\title{When to efficiently rebalance a portfolio}
\author{Masayuki Ando and Masaaki Fukasawa\footnote{The corresponding author.  Email: fukasawa@sigmath.es.osaka-u.ac.jp}\\
{\small Graduate School of Engineering Science, Osaka University, 560-8531 Japan}}

\date{}

\begin{document}

\maketitle

\begin{abstract}
    A constant weight asset allocation is a popular investment strategy and is optimal under a suitable continuous model. We study the tracking error for the target continuous rebalancing strategy by a feasible discrete-in-time rebalancing under a general multi-dimensional Brownian semimartingale model of asset prices. In a high-frequency asymptotic framework, we derive an asymptotically efficient sequence of simple predictable strategies.\\

    {\it Keywords.} Discretization of stochastic integrals, Asymptotic analysis, Constant weight asset allocation, Impulse control, Pearson's inequality. 
\end{abstract}

\section{Introduction}
Consider a multi-dimensional risky asset $S = (S^1,\dots, S^d)^\top$ and a risk-free asset $S^0$ with
\begin{equation}\label{BSM}
\frac{\mathrm{d}S^i_t}{S^i_t} = \mu_t^i \mathrm{d}t + \sum_{j=1}^m \sigma_t^{ij} \mathrm{d}W^j_t,
\ \ 
 \frac{\mathrm{d}S^0_t}{S^0_t}  = \mu^0_t \mathrm{d}t
\end{equation}
where $(W^1,\dots,W^m)$ is an $m$-dimensional standard Brownian motion, and
$\mu^i$ and  $\sigma^{ij}$ are locally bounded
adapted processes with
\begin{equation*}
\Sigma_t = [\Sigma^{ik}_t],\ \     \Sigma^{ik}_t := \sum_{j=1}^m \sigma^{ij}_t \sigma^{jk}_t
\end{equation*}
being positive definite for all $t\geq 0$. 
For any $d+1$ dimensional locally bounded adapted process $(\pi^0,\pi^1,\dots, \pi^d)$ with
$\sum_{i=0}^d \pi^i = 1$, the equation
\begin{equation}\label{eqV}
\frac{\mathrm{d} V_t}{V_t} = \sum_{i=0}^d \pi^i_t \frac{\mathrm{d}S^i_t}{S^i_t} - c_t \mathrm{d}t
\end{equation}
describes the dynamics of the wealth process $V$ associated with a self-financing strategy and a consumption plan $c$ under the admissibility constraint $V>0$.
The ratio of the wealth invested in $S^i$ to the 
total wealth $V$ is $\pi^i$. A constant weight asset allocation refers to such a strategy that each $\pi^i$ is kept constant. 

We assume that 
a continuous-time strategy $\pi = (\pi^1,\dots,\pi^d)^\top$ which is positive and  absolutely continuous with locally bounded derivative
is given to follow. 
For a given strategy $\pi$, $\pi^0$ is always set to be
$\pi^0 = 1 - \sum_{i=1}^d \pi^i$.
Recall that such $\pi$ appears as the growth optimal portfolio strategy when
\begin{equation*}
    \theta:= \Sigma^{-1}(\mu-r)
\end{equation*}
is positive and absolutely continuous with locally bounded derivative,
where $\mu = (\mu^1,\dots,\mu^d)^\top$ 
and $r = \mu^0(1,\dots,1)^\top$.
Indeed, if $c=0$,
\begin{equation*}
    \log V_T = \log V_0 + \int_0^T \frac{\mathrm{d}V_t}{V_t} -\frac{1}{2}\int_0^T \pi_t^\top \Sigma_t \pi_t \, \mathrm{d}t
\end{equation*}
and so, under a suitable admissibility condition,
\begin{equation*}
\mathsf{E}\left[\log \frac{V_T}{V_0}\right] = 
\int_0^T \mathsf{E}\left[ 
-\frac{1}{2}
(\pi_t -\theta_t)^\top \Sigma_t (\pi_t-\theta_t) + \frac{1}{2} \theta_t^\top \Sigma_t \theta_t 
\right]\, \mathrm{d}t,
\end{equation*}
which is maximized by $\pi = \theta$.
Under the Black-Scholes dynamics, where $\mu$, $r$ and $\Sigma$ are deterministic, 
the optimal strategy of the consumption and investment problem is known to be proportional to
the deterministic function $\theta$
 under power utilities \cite{KS}, or more generally, the Epstein-Zin stochastic differential utilities \cite{Kraft}, 
or even under relative performance criteria~\cite{LZ}.
The optimal strategies $\pi$ are then
absolutely continuous with locally bounded derivative
if so are  $\mu$, $r$ and $\Sigma$.
In particular, they are constant weight allocation strategies if $\mu$, $r$ and $\Sigma$ are constant.
Also under model uncertainty, the superiority of an equal-weighted portfolio, also known as the $1/N$ portfolio, has been documented in the literature (e.g., \cite{DGU}).
Beyond these theoretical frameworks, a constant weight asset allocation has been popular in the asset management industry, dated back to Talmud (1200 BC - 500 AD)~\cite{Gibson}.
In this paper,
 we assume a strategy $(\pi,c)$ to be given for whatever reason and consider how to implement it  under a general Brownian semimartingale model \eqref{BSM} and \eqref{eqV}.
 
Denote by $H = (H^1,\dots,H^d)^\top$, $H^i := V\pi^i/S^i$,
the numbers of shares  associated with the asset allocation strategy $\pi$. Notice that $H$ is not of finite variation even though $\pi$ is so. Indeed, we see in Section~2 that the quadratic covariation of $H$ is nondegenerate.
Now the question is how to implement $H$ in reality, where a continuous adjustment of portfolio is infeasible.
Asset re-allocations have to be discrete in time and should be as less frequent as possible to avoid various kind of costs. Then the question is when and how to rebalance a portfolio efficiently.

Finding an efficient discrete-in-time rebalancing strategy amounts to finding an efficient approximation to a stochastic integral by one with a simple predictable integrand.
In the case of $d=1$, an asymptotically efficient 
sequence of simple predictable approximations was derived in
\cite{F11a,F11b, F14}.
An extension to the multi-dimensional case in a hedging context was given by \cite{GL}, which however does not cover investment strategies such as constant weight asset allocations.
In this paper, we give an extension to this missing direction.
Further, in contrast to \cite{GL}, we do not restrict candidate strategies to discretization schemes but discuss asymptotic efficiency in a broader class of simple predictable strategies.
From a mathematical point of view, this extension involves a novel inequality for centered moments of a general random vector that generalizes Pearson's inequality for one-dimensional kurtosis and skewness.

For the multi-dimensional Black-Scholes model,
an asymptotic analysis of
the optimal consumption investment problem under
fixed transaction costs was given in \cite{Alt}.
Under the fixed transaction costs, the number of rebalancing penalizes the total wealth. The asymptotic solution of \cite{Alt} is a discretization of the Merton portfolio, a constant weight strategy which is optimal in the frictionless market, by a sequence of stopping times.
Although our optimization problem is different from \cite{Alt}, our solution has a similar structure to
that of \cite{Alt}, obtained by solving the same algebraic Riccati equation.

In Section~2, we compute the quadratic covariation $\langle H, H \rangle$ of $H$ 
when $\pi$ is positive and of finite variation. We observe a structural condition between $\langle H,H\rangle$ and $\langle S, S\rangle$ is met.
In Section~3, we state our main result relying on this structural condition under a more abstract framework of continuous semimartingales than \eqref{BSM} and \eqref{eqV}.
In Section~4, we derive an asymptotically efficient strategy and discuss the efficiency loss of the equidistant discretization. 
In Section~\ref{Proof}, we give the proof of the main theorem stated in Section~2.
In Section~\ref{KS}, we prove an
 inequality for centered moments of a general random vector that generalizes Pearson's inequality for one-dimensional kurtosis and skewness.

\section{The structure of the continuous strategy}
Here we compute the quadratic covariations of the process $H = (H^1,\dots,H^d)^\top$,
which plays a key role in our analysis in the next section, 
under the assumption that the asset allocation strategy 
$\pi$ is positive and of finite variation with \eqref{eqV}.

\begin{lem}\label{lem1}
Assume $\pi^i$ to be a positive continuous process of finite variation for all $i=1,\dots, d$.
Under \eqref{BSM} and \eqref{eqV},
    \begin{equation}\label{Hqv}
\mathrm{d}\langle H, H \rangle_t = (U_t)^\top U_t \mathrm{d}t,
\end{equation}
where
\begin{equation*}
U^i_t = H_t^i\Sigma_t^{1/2}(\pi_t-e_i),
\ \ U = (U^1,\dots, U^d)
\end{equation*}
and  $\{e_i\}_{i=1}^d$ is the standard basis of $\mathbb{R}^d$.
Further, if in addition 
$\pi^0_t  \neq 0 $ for all $t\geq 0$, then
$\det U_t \neq 0$ and
\begin{equation}\label{Sqv}
K^\top_t \mathrm{d}\langle H, H \rangle_t K_t = \mathrm{d}\langle S, S \rangle_t,
\end{equation}
where
$K_t = (U_t^\top U_t)^{-1/2}\Sigma_t^{1/2}$.
\end{lem}
{\it Proof: }
Recall that $H^i = V\pi^i/S^i$, so that
\begin{equation*}
\mathrm{d}\langle \log H^i, \log H^j \rangle_t
= \mathrm{d}\langle \log V - \log S^i, \log V - \log S^j \rangle_t
= (\pi_t-e_i)^\top\Sigma_t (\pi_t - e_j) \mathrm{d}t,
\end{equation*}
Therefore, 
\begin{equation*}
\mathrm{d}\langle H^i, H^j \rangle_t = (U^i_t)^\top U^j_t \mathrm{d}t,
\end{equation*}
which implies \eqref{Hqv}.

Now let us see that $\det U_t \neq 0$ if
$\pi_t^i  \neq 0 $ for all $i=0,1,\dots, d$.
We omit the dependence $t$ for brevity.
We are going to show $U^i$, $i=1,\dots, d$ are linearly independent. If there exists $a = (a^1,\dots,a^d)^\top \neq 0$ such that $Ua = 0$, then denoting $\tilde{H} = (\tilde{H}^1,\dots, \tilde{H}^d) = (H^1a^1,\dots, H^da^d)$,
\begin{equation*}
\left(\sum_{i=1}^d \tilde{H}^i\right)\Sigma^{1/2}\pi = \Sigma^{1/2}\tilde{H}^\top.
\end{equation*}
Since $\Sigma$ is positive definite, the row vectors of $\Sigma^{1/2}$ are linearly independent and so,  $\sum_{i=1}^d \tilde{H}^i \neq 0$. 
Then, we have
\begin{equation*}
\pi = \left(\sum_{i=1}^d \tilde{H}^i\right)^{-1} \tilde{H}^\top,
\end{equation*}
which contradicts $\pi^0 = 1 - \sum_{i=1}^d \pi^i \neq 0$.
\hfill{$\square$}
\\


\section{The main result}
Here we give a mathematical formulation of the problem and then state our main result.
Let 
$(\Omega,\mathcal{F},\mathsf{P},\{\mathcal{F}_t\}_{t\in [0,1]})$ be a filtered probability space satisfying the usual assumptions.
A simple predictable process is a stochastic process of the form
\begin{equation*}
 X = \sum_{i=0}^{\infty} \xi_i 1_{((\tau_i,\tau_{i+1}]]},
\end{equation*}
where $\{\tau_i\}_{i \geq 0} $ is a
nondecreasing sequence of stopping times taking values in $[0,1]$ and
$\xi_i$ is an $\mathcal{F}_{\tau_i}$ measurable $d$-dimensional
random variable. 
For $X$
of the above form and for
a $d$-dimensional continuous semimartingale $S$,
the stochastic integral $X\cdot S$ is defined by
\begin{equation*}
(X \cdot S)_t
= \sum_{i=0}^\infty \xi_i^\top (S_{\tau_{i+1} \wedge t} - S_{\tau_i \wedge t}).
\end{equation*}
For given $d$-dimensional continuous semimartingales $H$ and $S$,
we are interested in an efficient approximation to
$H\cdot S $ by a sequence $X^n\cdot S$, where $X^n$ are
simple predictable processes.

Denote by $\mathcal{M}_d$ and $\mathcal{S}_d$ respectively the sets of $d\times d$ regular matrices and positive definite matrices.
\begin{assumption}\label{ass1}
 There exist an $\mathcal{S}_d$-valued continuous adapted process $J$,  an $\mathcal{M}_d$-valued continuous
 adapted process $K$,
 and 
 a continuous nondecreasing adapted process $A$ such
that
\begin{equation*}
 \mathrm{d}\langle H, H \rangle =  J \, \mathrm{d} A,\ \
\mathrm{d}\langle S, S \rangle = K^\top J K \, \mathrm{d}A.
\end{equation*} 
  The finite variation part of $H$ is absolutely continuous with respect to $A$ and the associated Radon-Nikodym derivative is  locally bounded.
\end{assumption}
Under \eqref{BSM} with \eqref{eqV},
by Lemma~\ref{lem1}, for an asset allocation strategy $\pi$ being continuous and of finite variation with
$\pi^i_t \neq 0$ for all $t \in [0,1]$ and $i=0,1,\dots, d$,
Assumption~\ref{ass1} is satisfied with
$A_t =t$,
$J = U^\top U$, $K = (U^\top U)^{-1/2}\Sigma^{1/2}$,
where $U = (U^1,\dots, U^d)$, $U^i_t = H_t^i\Sigma_t^{1/2}(\pi_t-e_i)$.
The last condition on the finite variation part is satisfied 
if $\pi$ is absolutely continuous with locally bounded derivative
and if the consumption plan $c$ is also locally bounded.
\\

For positive continuous adapted processes $Q$ and $N$ fixed
and for  a simple predictable process $X$,
we introduce the cost functionals $Q[X]$ and $N[X]$ respectively of approximation error and of approximation effort as 
\begin{equation*}
 Q[X] = \int_0^1 Q_t \, \mathrm{d}\langle \, H\cdot S - X\cdot S \, \rangle_t,\ \ N[X] = \sum_{t \in (0,1)}N_t1_{\{|\Delta X_t| \neq 0\}}.
\end{equation*}
In particular, if $N=1$ then $N[X]$ counts the number of jumps of $X$, that is, the number of rebalancing in our financial context, and if $Q$ is the density process of 
an equivalent martingale measure $\mathsf{Q}$ for $S$ then $\mathsf{E}[Q[X]] = \mathsf{E}_\mathsf{Q}[(H\cdot S - X \cdot S)_1^2]$.
Note that the expected approximation error $\mathsf{E}[Q[X]]$ can be arbitrarily made small by taking $X$ sufficiently close to $H$, while it inevitably makes the expected approximation effort $\mathsf{E}[N[X]]$ large because $H$ has a nondegenerate quadratic variation.
We then seek an efficient frontier for the  trade-off between $\mathsf{E}[Q[X]]$ and $\mathsf{E}[N[X]]$. We take an asymptotic approach to have an explicit solution.

\begin{defn}
We say  a sequence of simple predictable processes $X^n$ is admissible if
\begin{enumerate}
\item $X^n$ is locally bounded for each $n$,
    \item $\displaystyle \sup_{t \in [0,1]}|X^n_t-H_t| \to 0$ in probability as $n\to \infty$,  
    \item $\mathsf{E}[Q[X^n]] < \infty$ and
    $\displaystyle \frac{Q[X^n]}{\mathsf{E}[Q[X^n]]}$ is uniformly integrable.
\end{enumerate}
\end{defn}
Now we state our main result, of which the proof is deferred to Section~\ref{Proof}.
\begin{thm}\label{thm1}
Let $H$ and $S$ be $d$-dimensional continuous semimartingales satisfying Assumption~\ref{ass1}, and let $Q$ and $N$ be 
 positive continuous adapted processes.
Then,
    for any admissible sequence $X^n$,
\begin{equation}\label{main2}
\varliminf_{n\to \infty} \mathsf{E}[N[X^n]] \mathsf{E}[Q[X^n]] \geq \mathsf{E}\left[ 
\int_0^1 N_t^{1/2}Q_t^{1/2}\mathrm{tr}(L_tJ_t)\mathrm{d}A_t \right]^2,
\end{equation}
where $L = \ell (J,K)$ and $\ell $ is the solution map given in Lemma~\ref{lem2}.
\end{thm}

\begin{lem}\label{lem2}

For any $J \in \mathcal{S}_d$ and $K \in \mathcal{M}_d$, there exists a unique
 $L = \ell (J,K) \in \mathcal{S}_d$ such that
\begin{equation*}
 2 \mathrm{tr}(LJ)L + 4LJL = K^\top JK.
\end{equation*}
Further, the map $\ell $ is continuous on $\mathcal{M}_d\times \mathcal{S}_d$.
\end{lem}
Lemma~\ref{lem2} is a straightforward extension of Lemma~3.1 of \cite{GL}, and so the proof is omitted.
This algebraic Riccati equation first appeared in \cite{Atk} to describe an approximate solution to the variational inequality for an optimal consumption investment problem under the Black-Scholes model with fixed-type transaction costs. The existence of the solution with an efficient computational algorithm was given in \cite{Atk}. 
The same equation naturally appeared in \cite{Alt}.

\section{Efficient and inefficient strategies}
\subsection{An asymptotically efficient sequence}
Here we show that the sequence
\begin{equation}\label{Xn}
 X^n = \sum_{i=0}^{\infty} \xi^n_i 1_{((\tau^n_i,\tau^n_{i+1}]]}
\end{equation}
defined by
\begin{equation}\label{eff}
    \xi^n_j = H_{\tau^n_j}, \ \ \tau^n_{j+1} = \inf\left\{ t > \tau^n_j \ ; \ (H_t-\xi^n_j)^\top L_{\tau^n_j}(H_t-\xi^n_j) = \epsilon_n Q_{\tau^n_j}^{-1/2}N_{\tau^n_j}^{1/2} \right\}
\end{equation}
and $\tau^n_0 = 0$
with a deterministic positive sequence $\epsilon_n$ with $\epsilon_n \to 0$ as $n\to \infty$ is asymptotically efficient.
\begin{thm}
Let $H$ and $S$ be $d$-dimensional continuous semimartingales satisfying Assumption~\ref{ass1}, and let $Q$ and $N$ be 
 positive continuous adapted processes. Then,
    \begin{equation}\label{qc}
    \epsilon_n^{-1}Q[X^n] \to \int_0^1 N_t^{1/2}Q_t^{1/2}\mathrm{tr}(L_tJ_t)\mathrm{d}A_t
    \end{equation}
    and
         \begin{equation}\label{nc}
              \epsilon_nN[X^n] \to \int_0^1 N_t^{1/2}Q_t^{1/2}\mathrm{tr}(L_tJ_t)\mathrm{d}A_t
         \end{equation}  
    in probability as $n\to \infty$.
\end{thm}
{\it Proof: }
By It\^o's formula,
\begin{equation}\label{Ito}
\begin{split}
      & \left( (H_{\tau^n_{j+1}}  - \xi^n_j)^\top L_{\tau^n_j} (H_{\tau^n_{j+1}} - \xi^n_j)\right)^2 
       \\ & =  \,  \left( (H_{\tau^n_{j}}- \xi^n_j)^\top L_{\tau^n_j} (H_{\tau^n_{j}} - \xi^n_j)\right)^2 \\
       &\ + 4\int_{\tau^n_j}^{\tau^n_{j+1}} 
       (H_t-\xi^n_j)^\top L_{\tau^n_j} (H_t-\xi^n_j)
       (H_t-\xi^n_j)^\top L_{\tau^n_j} \, \mathrm{d}H_t \\
       &\ + \int_{\tau^n_j}^{\tau^n_{j+1}}
       (H_t-\xi^n_j)^\top\left(2 \mathrm{tr}(L_{\tau^n_j}J_t) L_{\tau^n_j} + 4 L_{\tau^n_j}J_t L_{\tau^n_j} \right)(H_t-\xi^n_j)\,\mathrm{d}A_t.
\end{split}
\end{equation}
Therefore for \eqref{eff},
\begin{equation*}
\begin{split}
    \epsilon_n^{-1} Q[X^n] &= \epsilon_n^{-1}\int_0^1  (H_t-X^n_t)^\top K_t^\top J_t K_t (H_t-X^n_t) Q_t\, \mathrm{d}A_t \\
     &= \sum_{j=0}^{\infty} Q_{\tau^n_j}^{1/2}N_{\tau^n_j}^{1/2} (H_{\tau^n_{j+1}}  - H_{\tau^n_j})^\top L_{\tau^n_j} (H_{\tau^n_{j+1}} - H_{\tau^n_j})
     + E^n_1 + E^n_2,
\end{split}
\end{equation*}
where
\begin{equation*}
\begin{split}
      & E^n_1 = \epsilon_n^{-1}
     \sum_{j=0}^{\infty} 
     \int_{\tau^n_j}^{\tau^n_{j+1}}  (H_t-H_{\tau^n_j})^\top E^{n,j}_t
    (H_t-H_{\tau^n_j}) \, \mathrm{d}A_t,\\
    &E^n_2 = 4\epsilon_n^{-1}
    \sum_{j=0}^{\infty} 
     \int_{\tau^n_j}^{\tau^n_{j+1}} 
       (H_t-H_{\tau^n_j})^\top L_{\tau^n_j}(H_t-H_{\tau^n_j})
       (H_t-H_{\tau^n_j})^\top L_{\tau^n_j} \, \mathrm{d}H_t, \\
   &E^{n,j}_t = K_t^\top J_t K_t Q_t - 
    \left(2 \mathrm{tr}(L_{\tau^n_j}J_t) L_{\tau^n_j} + 4 L_{\tau^n_j}J_t L_{\tau^n_j} \right)Q_{\tau^n_j}.
\end{split}
\end{equation*}
Since $L = \ell (J,K)$ and $\ell$ is continuous by Lemma~\ref{lem1}, we have
\begin{equation*}
    \sup_{t \in [0,1], j \geq 0} \left|E^{n,j}_{t \wedge \tau^n_{j+1}}\right| \to 0
\end{equation*}
in probability. Note also that 
\begin{equation}\label{fin}
    \sup_{t \in [0,1], j \geq 0} \epsilon_n^{-1}
 (H_{t \wedge \tau^n_{j+1}}-H_{\tau^n_j})^\top 
    (H_{t \wedge \tau^n_{j+1}}-H_{\tau^n_j}) < \infty
\end{equation}
 under \eqref{eff}. These imply that $E^n_1\to 0$ in probability.
 We also have $E^n_2\to 0$ in probability because
 \begin{equation*}
     \epsilon_n^{-2} \int_0^1 
      ((H_t-X^n_t)^\top (H_t-X^n_t))^3 \mathrm{tr} (J_t)
        \, \mathrm{d}A_t \to 0
 \end{equation*}
 in probability by \eqref{fin} again, with the aid of the Lenglart inequality.
 We then conclude \eqref{qc}. 
 
 To see \eqref{nc}, observe that
 \begin{equation*}
     \epsilon_nN[X^n] = \sum_{j=0}^\infty N_{\tau^n_{j+1}}
     Q_{\tau^n_j}^{1/2}N_{\tau^n_j}^{-1/2} 
     (H_{\tau^n_{j+1}}  - H_{\tau^n_{j}})^\top L_{\tau^n_j} (H_{\tau^n_{j+1}} - H_{\tau^n_{j}}).
 \end{equation*}
 under \eqref{eff}.
\hfill{$\square$}\\

It is not difficult to see that $X^n$ is admissible and  attains the equality in \eqref{main2} under suitable additional conditions on $J$, $K$, $Q$ and $N$, $X^n$.

\begin{rem}\upshape
Under \eqref{BSM} and \eqref{eqV},
\begin{equation*}
    V_t = V_0 \exp\left\{
\sum_{i=0}^d \pi^i_u \frac{\mathrm{d}S^i_u}{S^i_u}
- \int_0^t \left(c_u + \frac{1}{2}\pi_u^\top\Sigma_u\pi_u \right) \, \mathrm{d}u
    \right\}.
\end{equation*}
Notice that
\begin{equation*}
    H^i_{\tau^n_j} = \frac{\pi^i_{\tau^n_j} }{S^i_{\tau^n_j}}
    V_{\tau^n_j} = \hat{\xi}^{n,i}_j + 
     \frac{\pi^i_{\tau^n_j} }{S^i_{\tau^n_j}}
    (V_{\tau^n_j} - V^n_{\tau^n_j}),
\end{equation*}
where $\hat{\xi}^{n,i}_j = \pi^i_{\tau^n_j} V^n_{\tau^n_j}/S^i_{\tau^n_j}$ is the number of share to invest $\pi^i_{\tau^n_j}$ portion of the total wealth
\begin{equation}\label{Vn}
    V^n_{\tau^n_j} = V_0 + \int_0^{\tau^n_j}
     \frac{V^n_t-  (X^n_t)^\top S_t }{S^0_t} \mathrm{d}S^0_t+  \int_0^{\tau^n_j} (X^n_t)^\top \mathrm{d}S_t - \int_0^{\tau^n_j} c_t \, \mathrm{d}t
\end{equation}
in $S^i$ at time $\tau^n_j$.
\end{rem}

\subsection{The equidistant discretization}
Here we compute the efficiency loss for the equidistant discretization strategy
\begin{equation*}
    \xi^n_j = H_{\tau^n_j}, \ \ \tau^n_j = \frac{j}{n}
\end{equation*}
under the additional assumption that $A_t =t$.

\begin{thm}
Let $H$ and $S$ be $d$-dimensional continuous semimartingales satisfying Assumption~\ref{ass1} with $A_t = t$ for $t \in [0,1]$ and $J$ and $K$ being $h$-H\"older continuous for some $h>0$. Let $Q$ and $N$ be 
 positive  $h$-H\"older continuous adapted processes. Then,
    \begin{equation}\label{qc2}
    n Q[X^n] \to \int_0^1 Q_t (\mathrm{tr}(L_tJ_t)^2 + 2\mathrm{tr}(L_tJ_tL_tJ_t)) \, \mathrm{d}t
    \end{equation}
    and
         \begin{equation}\label{nc2}
              \epsilon_nN[X^n] \to \int_0^1 N_t \, \mathrm{d}t
         \end{equation}  
    in probability as $n\to \infty$.
\end{thm}
{\it Proof: }
Under the additional assumption of $A_t =t$, we know that $S$ and $H$ are Brownian semimartingales and in particular their sample paths are $1/2-\epsilon$ H\"older continuous almost surely for any $\epsilon>0$. Therefore, using \eqref{Ito},
we have
\begin{equation*}
\begin{split}
   n Q[X^n] &= n\int_0^1  (H_t-X^n_t)^\top K_t^\top J_t K_t (H_t-X^n_t) Q_t\, \mathrm{d}A_t \\
     &= n\sum_{j=0}^{\infty} Q_{\tau^n_j} \left( (H_{\tau^n_{j+1}}  - H_{\tau^n_{j}})^\top L_{\tau^n_j} (H_{\tau^n_{j+1}} - H_{\tau^n_{j}})\right)^2
     + E^n
\end{split}
\end{equation*}
with $E^n$ converging to $0$ in probability. 
On the other hand,  for $L$, $J \in \mathcal{S}_d$ and 
a Gaussian random vector $X = (X_1,\dots,X_d)\sim \mathcal{N}(0,J)$, we have
\begin{equation*}
\begin{split}
     &\mathsf{E}\left[\left(\sum_{i,j=1}^d X_iX_jL^{ij}\right)^2\right] = 
    \sum_{i,j,k,l=1}^d \mathsf{E}[X_iX_jX_kX_l]L^{ij}L^{kl} \\
    &= 
    \sum_{i,j,k,l=1}^d (\mathsf{E}[X_iX_j]\mathsf{E}[X_kX_l]
    + \mathsf{E}[X_iX_k]\mathsf{E}[X_jX_l] + \mathsf{E}[X_iX_l]\mathsf{E}[X_kX_j])L^{ij}L^{kl} \\
    &= \mathrm{tr}(LJ)^2 + 2 \mathrm{tr}(LJLJ)
\end{split}
\end{equation*}
by Isserlis' theorem. Then by a standard argument in the high-frequency data analysis (see e.g., \cite{AJ} or \cite{JP}), we obtain
\begin{equation*}
    n\sum_{j=0}^{\infty} Q_{\tau^n_j} \left( (H_{\tau^n_{j+1}}  - H_{\tau^n_{j}})^\top L_{\tau^n_j} (H_{\tau^n_{j+1}} - H_{\tau^n_{j}})\right)^2
    \to \int_0^1 Q_t (\mathrm{tr}(L_tJ_t)^2 + 2\mathrm{tr}(L_tJ_tL_tJ_t)) \, \mathrm{d}t
\end{equation*}
in probability. Thus we conclude \eqref{qc2}, while \eqref{nc2} is trivial.
\hfill{$\square$}
\\

The efficiency loss for the equidistant discretization can be decomposed into two parts. First,
\begin{equation*}
    \mathsf{E}\left[ 
\int_0^1 N_t^{1/2}Q_t^{1/2}\mathrm{tr}(L_tJ_t)\mathrm{d}t \right]^2
\leq  \mathsf{E}\left[ \int_0^1 N_t \, \mathrm{d}t \right]
 \mathsf{E}\left[ \int_0^1 Q_t \mathrm{tr}(L_tJ_t)^2\, \mathrm{d}t \right]
\end{equation*}
by the Cauchy-Schwarz inequality. Second,
\begin{equation*}
    \mathsf{E}\left[ \int_0^1 Q_t \mathrm{tr}(L_tJ_t)^2\, \mathrm{d}t \right]
    \leq 
     \mathsf{E}\left[ \int_0^1 Q_t \mathrm{tr}(L_tJ_t)^2
     \left(1 + \frac{2\mathrm{tr}(L_tJ_tL_tJ_t)}{\mathrm{tr}(L_tJ_t)^2} \right)
     \, \mathrm{d}t \right].
\end{equation*}
The loss from the first inequality is due to that the equidistant scheme does not take the time varying nature of volatility into account.
The loss from the second inequality is due to the use of deterministic time (or more generally, strongly predictable time; see \cite{AJ}). Indeed, the factor 
$1 + 2/d$ for the case of $LJ$ being the identity matrix coincides with the ratio of the asymptotic variance of the equidistant Euler-Maruyama scheme for discretizing stochastic differential equations to that of its hitting time counterpart given by \cite{FO}.

\section{Proof of Theorem~\ref{thm1}}\label{Proof}
It suffices to consider a case where $\mathsf{E}[N[X^n]]\mathsf{E}[Q[X^n]]$ converges.
Then, since $Q[X^n]/\mathsf{E}[Q[X^n]]$ is uniformly  integrable  so is $\mathsf{E}[N[X^n]]Q[X^n]$.
By localization, we can also assume without loss of generality that
all the locally bounded processes are bounded, and that 
all the positive continuous processes, including
the smallest eigenvalues of $\mathcal{S}^d$ valued continuous processes $J$ and $K^\top JK$,
are bounded away from $0$.
Let
\begin{equation*}
    X^n = 
    \sum_{j=0}^\infty \xi^n_j 1_{((\tau^n_j,\tau^n_{j+1}]]}
\end{equation*}
and 
\begin{equation*}
    Y^n = \sum_{j=0}^\infty Y_{\tau^n_j}
    1_{((\tau^n_j,\tau^n_{j+1}]]}
\end{equation*}
for $Y = J, K, L$ and $Q$. 
Since $\sup_{0\leq t \leq 1} |X^n_t-X_t| \to 0$ in probability, we have that
$\sup_{j \geq 0}|\tau^n_{j+1}-\tau^n_j| \to 0$ in probability and as a result, 
$\sup_{0\leq t \leq 1} |Y^n_t-Y_t| \to 0$ in probability
for $Y = J, K, L$ and $Q$. 
 We refer to \cite{F14} for more technical details on these observations in the one dimensional case; the proofs are trivially extended to the multi-dimensional case. 
 
By \eqref{Ito}, we have
\begin{equation*}
\begin{split}
        & \left( (H_{\tau^n_{j+1}}  - \xi^n_j)^\top L_{\tau^n_j} (H_{\tau^n_{j+1}} - \xi^n_j)\right)^2 
         \\  & =   \left( (H_{\tau^n_{j}}- \xi^n_j)^\top L_{\tau^n_j} (H_{\tau^n_{j}} - \xi^n_j)\right)^2 \\
       &\ \  + 4\int_{\tau^n_j}^{\tau^n_{j+1}} 
       (H_t-X^n_t)^\top L^n_t (H_t-X^n_t)
       (H_t-X^n_t)^\top L^n_t \, \mathrm{d}H_t \\
       &\ \ + \int_{\tau^n_j}^{\tau^n_{j+1}}
       (H_t-X^n_t)^\top\left(2 \mathrm{tr}(L^n_tJ_t) L^n_t + 4 L^n_tJ_t L^n_t \right)(H_t-X^n_t)\,\mathrm{d}A_t
\end{split}
\end{equation*}
and so,
\begin{equation*}
\begin{split}
     Q[X^n] &= \int_0^1  (H_t-X^n_t)^\top K_t^\top J_t K_t (H_t-X^n_t) Q_t\, \mathrm{d}A_t \\
     &= \sum_{j=0}^{\infty} G^n_j Q_{\tau^n_j}\left(((\Delta^n_j + \delta^n_j)^\top (\Delta^n_j + \delta^n_j))^2 - ((\delta^n_j)^\top \delta^n_j)^2\right)
     + E^n_1 + E^n_2,
\end{split}
\end{equation*}
where 
\begin{equation*}
\begin{split}
& \Delta^n_j = L_{\tau^n_j}^{1/2}  (H_{\tau^n_{j+1}}  - H_{\tau^n_j}),\\
& \delta^n_j = L_{\tau^n_j}^{1/2}  (H_{\tau^n_j}  - \xi^n_j),\\
      & E^n_1 = 
    \int_0^1  (H_t-X^n_t)^\top \left(K_t^\top J_t K_t Q_t - 
    \left(2 \mathrm{tr}(L^n_tJ_t) L^n_t + 4 L^n_tJ_t L^n_t \right)Q^n_t G^n_t\right)
    (H_t-X^n_t) \, \mathrm{d}A_t,\\
    &E^n_2 = 4
    \sum_{j=0}^{\infty} G^n_j 
     \int_{\tau^n_j}^{\tau^n_{j+1}} 
       (H_t-X^n_t)^\top L^n_t (H_t-X^n_t)
       (H_t-X^n_t)^\top L^n_t \, \mathrm{d}H_t,
       \\
       & G^n = \sum_{j=0}^\infty G^n_j1_{((\tau^n_j,\tau^n_{j+1}]]}, \ \ G^n_j = 
       \exp\left\{ -\int_{\tau^n_j}^{\tau^n_{j+1}} G_t^\top J_t^{-1} \mathrm{d} M_t - \frac{1}{2}\int_{\tau^n_j}^{\tau^n_{j+1}} G_t^\top J_t^{-1} G_t \, \mathrm{d}A_t\right\}
\end{split}
\end{equation*}
and $M$ and $G$ are respectively the local martingale part of $H$ and the Radon-Nikodym derivative of the finite variation part of $H$ with respect to $A$.

Since $L = \ell (J,K)$ and $\ell$ is continuous by Lemma~\ref{lem1}, we have
\begin{equation*}
\begin{split}
&    \sup_{0\leq t \leq 1} \left|K_t^\top J_t K_t Q_t - 
    \left(2 \mathrm{tr}(L^n_tJ_t) L^n_t + 4 L^n_tJ_t L^n_t \right)Q^n_t \right|
    \\
    & =  \sup_{0\leq t \leq 1} \left|\left(2 \mathrm{tr}(L_tJ_t) L_t + 4 L_tJ_t L_t \right) Q_t - 
    \left(2 \mathrm{tr}(L^n_tJ_t) L^n_t + 4 L^n_tJ_t L^n_t \right)Q^n_t \right| \to 0
    \end{split}
\end{equation*}
in probability. Together with $\sup_{0 \leq t \leq 1}|G^n_t-1| \to 0$ and the uniform integrability of
$\mathsf{E}[N[X^n]]Q[X^n]$, we deduce
$\mathsf{E}[N[X^n]]\mathsf{E}[|E^n_1|] \to 0$.

Define probability measures $\mathsf{Q}^n_j$ by
\begin{equation*}
\frac{\mathrm{d}\mathsf{Q}^n_j}{\mathrm{d}\mathsf{P}} = G^n_j.
\end{equation*}
By the Girsanov-Maruyama theorem, $H_{\cdot \wedge \tau^n_{j+1}} - H_{\cdot \wedge \tau^n_j}$ is a martingale under  $\mathsf{Q}^n_j$ for each $j \geq 0$.
This implies $\mathsf{E}[E^n_2] = 0$ and 
\begin{equation*}
\begin{split}
    &
        \mathsf{E}\left[
     \sum_{j=0}^{\infty} G^n_j Q_{\tau^n_j}\left(((\Delta^n_j + \delta^n_j)^\top (\Delta^n_j + \delta^n_j))^2 - ((\delta^n_j)^\top \delta^n_j)^2\right)
\right]\\
&= 
 \mathsf{E}\left[
     \sum_{j=0}^{\infty}  Q_{\tau^n_j}
     \mathsf{E}_{\mathsf{Q}^n_j}\left[\left(((\Delta^n_j + \delta^n_j)^\top (\Delta^n_j + \delta^n_j))^2 - ((\delta^n_j)^\top \delta^n_j)^2\right) | \mathcal{F}_{\tau^n_j}\right]
\right].
\end{split}
\end{equation*}
Here we have used the fact that all the partial sums of the infinite series are uniformly bounded as shown by rewriting them as integrals using It\^o's formula.
Further by Lemma~\ref{lemKS} in Section~\ref{KS},
this expectation is lower  bounded by
\begin{equation*}
 \mathsf{E}\left[
     \sum_{j=0}^{\infty}  Q_{\tau^n_j}
     \mathsf{E}_{\mathsf{Q}^n_j}\left[(\Delta^n_j )^\top \Delta^n_j  | \mathcal{F}_{\tau^n_j}\right]^2
\right].
\end{equation*}
Thus,
\begin{equation*}
\begin{split}
      \lim_{n \to \infty}
    \mathsf{E}[N[X^n]]\mathsf{E}[Q[X^n]]
    &\geq    \varlimsup_{n \to \infty} \mathsf{E}[N[X^n]]
    \mathsf{E}\left[
     \sum_{j=0}^{\infty}  Q_{\tau^n_j}
     \mathsf{E}_{\mathsf{Q}^n_j}\left[(\Delta^n_j )^\top \Delta^n_j  | \mathcal{F}_{\tau^n_j}\right]^2 \right]\\
     &\geq
      \varlimsup_{n \to \infty}
    \mathsf{E}\left[
     \sum_{j=0}^{\infty}  N_{\tau^n_j}^{1/2}Q_{\tau^n_j}^{1/2}
     \mathsf{E}_{\mathsf{Q}^n_j}\left[(\Delta^n_j )^\top \Delta^n_j  | \mathcal{F}_{\tau^n_j}\right] \right]^2\\
     &=
      \varlimsup_{n \to \infty}
    \mathsf{E}\left[
     \sum_{j=0}^{\infty}  N_{\tau^n_j}^{1/2}Q_{\tau^n_j}^{1/2} G^n_j
    \int_{\tau^n_j}^{\tau^n_{j+1}} \mathrm{tr}(L_{\tau^n_j} J_t)\, \mathrm{d}A_t \right]^2\\
    &= \mathsf{E}\left[\int_0^1 N_t^{1/2}Q_t^{1/2}
    \mathrm{tr}(L_t J_t)\, \mathrm{d}A_t \right]^2
\end{split} 
\end{equation*}
with the aid of the Cauchy-Schwarz inequality. 

\section{Kurtosis-Skewness inequality}\label{KS}
Here we prove an inequality for centered fourth and third moments of a general random vector.
This is a version of multi-variate Pearson's inequality; see \cite{Mori, Ogasawara} for related preceding results.

\begin{lem}\label{lemKS}
Let $\Delta$ be a $d$-dimensional $L^4$ random variable with
$\mathsf{E}[\Delta] = 0$ and $\delta \in \mathbb{R}^d$.
Then,
\begin{equation*}
    \mathsf{E}[((\Delta + \delta)^\top(\Delta + \delta))^2] - (\delta^\top\delta)^2
    \geq \mathsf{E}[\Delta^\top\Delta]^2.
\end{equation*}
\end{lem}
{\it Proof:} We have
\begin{equation*}
\begin{split}
& \mathsf{E}[((\Delta + \delta)^\top(\Delta + \delta))^2] - (\delta^\top\delta)^2\\
&= \mathsf{E}[(\Delta^\top\Delta + 2\delta^\top\Delta + \delta^\top\delta)^2] -
 (\delta^\top\delta)^2 \\
&= \mathsf{E}[(\Delta^\top\Delta)^2] + 4 \delta^\top \mathsf{E}[\Delta\Delta^\top]\delta +
4\mathsf{E}[\delta^\top\Delta
 (\Delta^\top\Delta)] + 2 \delta^\top \delta \mathsf{E}[\Delta^\top\Delta].
\end{split}
\end{equation*}
Taking the gradient with respect to $\delta$, 
\begin{equation*}
2 (4\mathsf{E}[\Delta\Delta^\top]) + 2\mathsf{E}[\Delta^\top\Delta]) \delta + 
4\mathsf{E}[\Delta
 (\Delta^\top\Delta)] 
\end{equation*}
and so, the minimum is attained at
\begin{equation*}
 \delta = - (2\mathsf{E}[\Delta\Delta^\top] +
  \mathsf{E}[\Delta^\top\Delta])^{-1}
\mathsf{E}[\Delta
 (\Delta^\top\Delta)].
\end{equation*}
Substitute this to get
\begin{equation*}
\begin{split}
& \mathsf{E}[((\Delta + \delta)^\top(\Delta + \delta))^2] -
  (\delta^\top\delta)^2
\\&\geq \mathsf{E}[(\Delta^\top\Delta)^2] - \mathsf{E}[
 (\Delta^\top\Delta)\Delta^\top]
\left(\mathsf{E}[\Delta\Delta^\top] + \frac{1}{2}
  \mathsf{E}[\Delta^\top\Delta] I \right)^{-1}
\mathsf{E}[\Delta
 (\Delta^\top\Delta)].
\end{split}
\end{equation*}
The result then follows from the Lemma~\ref{mks}.
\hfill{$\square$}

\begin{lem}\label{mks}
    Let $\Delta$ be a $d$-dimensional $L^4$ random variable with
$\mathsf{E}[\Delta] = 0$ and $\delta \in \mathbb{R}^d$.
Then,
\begin{equation*}
 \mathsf{E}[(\Delta^\top\Delta)^2] - \mathsf{E}[
 (\Delta^\top\Delta)\Delta^\top]
\left(\mathsf{E}[\Delta\Delta^\top] + D \right)^{-1}
\mathsf{E}[\Delta
 (\Delta^\top\Delta)]
\geq \mathsf{E}[\Delta^\top\Delta]^2
\end{equation*}
for any $D \in \mathcal{S}_d$.
\end{lem}
{\it Proof: }
For any $\alpha \in \mathbb{R}$ and $\beta \in
\mathbb{R}^d$,
\begin{equation*}
 \mathsf{E}[(\alpha (\Delta^\top\Delta-\mathsf{E}[\Delta^\top\Delta]) +
  \beta^\top\Delta)^2] \geq 0
\end{equation*}
The left hand side is a quadratic form with respect to the symmetric
matrix
\begin{equation*}
 \begin{pmatrix}
\mathsf{E}[(\Delta^\top\Delta-\mathsf{E}[\Delta^\top\Delta])^2] & 
\mathsf{E}[\Delta^\top
 (\Delta^\top\Delta)] \\
\mathsf{E}[\Delta 
 (\Delta^\top\Delta)] & \mathsf{E}[\Delta \Delta^\top]
 \end{pmatrix}
\end{equation*}
and the above nonnegativity implies that the matrix is nonnegative
definite. Therefore the matrix
\begin{equation*}
 \begin{pmatrix}
\mathsf{E}[(\Delta^\top\Delta-\mathsf{E}[\Delta^\top\Delta])^2] & 
\mathsf{E}[\Delta^\top
 (\Delta^\top\Delta)] \\
\mathsf{E}[\Delta 
 (\Delta^\top\Delta)] & \mathsf{E}[\Delta \Delta^\top] + D
 \end{pmatrix}
\end{equation*}
is also nonnegative definite and so, has a nonnegative determinant.
By the determinant formula for block matrices, 
the determinant is computed as
\begin{equation*}
\begin{split}
& \left|
\mathsf{E}[\Delta \Delta^\top] + D
\right|\\
& \times  \left(\mathsf{E}[(\Delta^\top\Delta - \mathsf{E}[\Delta^\top\Delta])^2] - \mathsf{E}[
 (\Delta^\top\Delta)\Delta^\top]
\left(\mathsf{E}[\Delta\Delta^\top] + D\right)^{-1}
\mathsf{E}[\Delta
 (\Delta^\top\Delta)]
\right),
\end{split}
\end{equation*}
which implies the claim.
\hfill{$\square$}

\begin{rem}\upshape
As easily seen from the proof, the equality is attained in Lemma~\ref{mks} when
$\Delta^\top \Delta = \mathsf{E}[\Delta^\top\Delta]$, or equivalently, $\Delta$ is supported on a sphere.
We apply the inequality in Section~\ref{Proof} for 
$\Delta = L_{\tau^n_j}^{1/2}(X_{\tau^n_{j+1}}-X_{\tau^n_j})$, so we have 
$\Delta^\top \Delta = \mathsf{E}[\Delta^\top\Delta]$ when
$X_{\tau^n_{j+1}}-X_{\tau^n_j}$ is supported on an ellipsoid characterized by $L_{\tau^n_j}$.
This explains the construction of our efficient strategy in Section~4. 
\end{rem}

\noindent
{\bf Acknowledgements.}
The authors thank Jun Sekine and Mike Tehranchi for helpful discussions.

\end{document}